# Highly Efficient Year-Round Energy and Comfort Optimization of HVAC Systems in Electric City Buses ⋆

**Fabio Widmer** * **Andreas Ritter** * **Mathias Achermann** *
**Fabian Büeler** * **Joshua Bagajo** * **Christopher H. Onder** *

* *Institute for Dynamic Systems and Control, ETH Zürich, 8092 Zürich, Switzerland (e-mail: fawidmer@idsc.mavt.ethz.ch)*

**Abstract:** In this paper, we present a novel approach to perform highly efficient numerical simulations of the heating, ventilation, and air-conditioning (HVAC) system of an electric city bus. The models for this simulation are based on the assumption of a steady-state operation. We show two approaches to obtain the minimum energy requirement for a certain thermal comfort criterion under specific ambient conditions. Due to the computationally efficient approach developed, we can evaluate the model on a large dataset of 7500 scenarios in various ambient conditions to estimate the year-round performance of the system subject to different comfort requirements. Compared to a heating strategy based on positive temperature coefficient (PTC) elements, we can thus show that a heat pump (HP) can reduce the annual mean power consumption by up to 60%. Ceiling-mounted radiant heating elements complementing a PTC heating system can reduce the annual mean power consumption by up to 10%, while they cannot improve the energy efficiency when used in conjunction with a HP. Finally, a broad sensitivity study reveals the fact that improving the HP's heating-mode coefficient of performance (COP) manifests the largest leverage in terms of mean annual power consumption. Moreover, the annual energy expenditure for cooling are around eight times smaller than those for heating. The case study considered thus reveals that the advantages of improving the COP of the cooling mode are significantly lower.

## 1. INTRODUCTION

### 1.1 Motivation

Global warming is one of the greatest challenges currently faced by humanity. One of the necessary steps to deal with this challenge is to phase out the use of fossil fuels. This urge is reinforced by the current global energy crisis, initially caused by the COVID-19 pandemic and further fueled in Europe by the military conflict between Russia and Ukraine. According to Bibra et al. (2022), these effects are leading to a massive worldwide increase of the sales of battery electric vehicles (BEVs) .

However, although BEVs offer clear advantages with regard to their propulsion efficiency, their relatively high demand of energy for the heating, ventilation, and air-conditioning (HVAC) system is still an open issue. This is especially problematic as high HVAC consumption can significantly impact the achievable driving range, which is a sensitive issue for BEVs in general. For urban public transport vehicles, these effects are even more pronounced due to the poor insulation of the passenger cabin and the frequent door openings. For instance, Cigarini et al. (2021) show that in winter conditions the HVAC energy consumption of an electric city bus can amount to almost half the overall energy consumption.

Therefore, analyzing and optimizing the HVAC systems is a very important topic in research and industry, not only in public transport, but for all types of vehicles. For instance, various suppliers have joined forces to develop a prototype electric bus with a heat pump (HP), seat heaters, and phase change material (PCM) heating elements, as reported by Hanke (2018), in order to demonstrate the applicability of these concepts in practice. Additionally, the use of radiant heater (RH) panels or alternative refrigerants like $CO_2$ (R744) are investigated for passenger vehicles by Qi (2014) and Cvok et al. (2021), respectively. Evaluating and comparing the performance of such concepts requires a method to efficiently calculate the achievable trade-off between the energy consumption and the thermal comfort of a given HVAC concept on a set of scenarios that represent a typical year-round operation.

### 1.2 Literature

In the scientific context, such evaluations typically are based on numerical simulations, which allow a much faster evaluation of changes in HVAC concepts than physical experiments permit. Most of the work presented in the literature thus relies on dynamic models, which allow simulations of transient processes, e.g., the initial cabin conditioning (heat-up or cool-down) as studied by Ramsey et al. (2022). Such transient processes are particularly relevant for passenger vehicles, where the typical duration of a trip is not significantly longer than such transient phases. Also, dynamic models are required to analyze or compare control approaches, as presented by Schaut and Sawodny (2020), for instance. However, as these dynamic simulations are numerically expensive, typically, only single scenarios are considered, the results of which may or may not be representative for the year-round operation of a vehicle.

⋆ This work has been supported by the Swiss Federal Office of Energy (SFOE, contract number SI/501979-01) and the industrial partners Carrosserie HESS AG and Verkehrsbetriebe Zürich (VBZ).

### 1.3 Research Statement

In contrast to such dynamic models, we suggest the use of steady-state models for the analysis of HVAC systems in urban public transport vehicles such as city buses. The rationale for this proposal is a clear separation of timescales between the relevant dynamics, which are in the order of tens of minutes, and the disturbances, whose time constants are in the order of tens of seconds (door openings, changes of driving speed) or hours (ambient conditions like temperature). Throughout the period of one hour, the operation of the HVAC system of such vehicles can thus be considered to be in a steady state. Clearly, these assumptions are not valid for many other thermal systems, such as passenger cars during heat-up conditions or buildings, the thermal dynamics of which are much slower and are thus in the same order of magnitude as those of changes of ambient conditions.

A steady-state model, combined with an efficient evaluation algorithm, thus can be a viable tool to analyze the HVAC systems of city buses. By neglecting all system dynamics, the evaluation of the trade-off between the power consumption and the thermal comfort for specific ambient conditions is much faster, which allows the system to be exposed not only to a small selection of ambient conditions, but to a great number of scenarios representative of a year-round operation. To the best of our knowledge, such a holistic sensitivity analysis of the HVAC system of an electric city bus based on steady-state calculations has not yet been proposed in the literature.

### 1.4 Paper Structure

This text is structured as follows: Section 2 introduces the mathematical models. Section 3 presents two approaches to determine the smallest power consumption of the system necessary to satisfy the thermal comfort requirements. Section 4 introduces the scenarios defining the ambient conditions of the case study, which are then used to generate the results shown in Section 5. Finally, Section 6 concludes this work and lists potential future research directions.

## 2. MODEL

Figure 1 provides a graphical overview of all model components. It shows the four thermal reservoirs with their respective temperature variables, i.e., the RH temperature $T_{\mathrm{rh}}$, the cabin air temperature $T_{\mathrm{cab}}$, and the inside and outside shell surface temperatures $T_{\mathrm{s,i}}$ and $T_{\mathrm{s,o}}$. The ambient is assumed to be at a uniform temperature $T_{\infty}$.

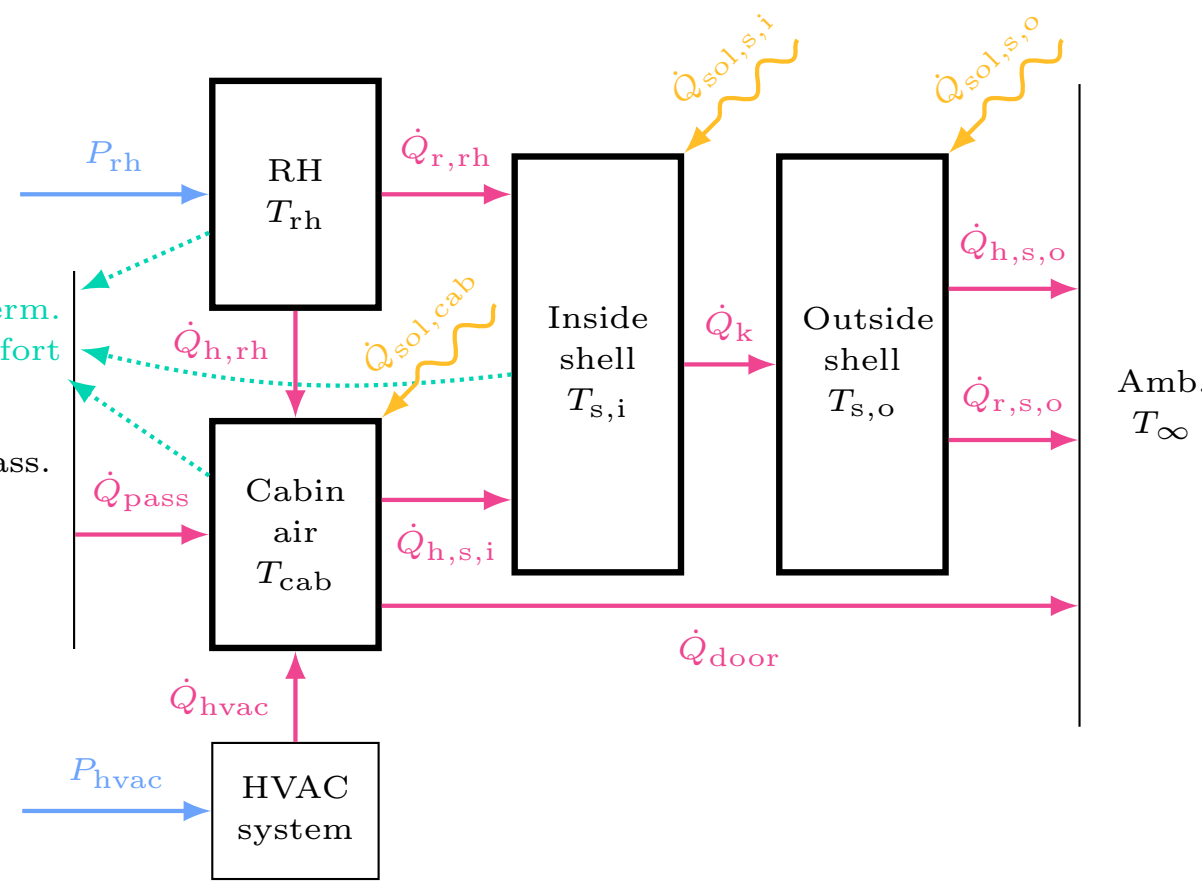

Fig. 1. Model overview including the thermal reservoirs (bold blocks) and the connecting flows. Red and yellow arrows denote heat flows, blue arrows denote electric power flows, while green dotted arrows denote influences on thermal comfort.

This section provides an introduction to the models needed to describe the system's behavior, starting with the models for the thermal and electric flows and followed by the models for determining thermal comfort.

### 2.1 Heat Flows

*Metabolic Heat:* The heat emitted by $N_{\mathrm{pass}}$ passengers is described by

$$\dot{Q}_{\mathrm{pass}} = N_{\mathrm{pass}} \cdot \dot{Q}_{\mathrm{met}}\,, \tag{1}$$

where $\dot{Q}_{\mathrm{met}} = 125\,\mathrm{W}$ represents the metabolic heat rate of the human organism, assuming the passengers are seated according to EN ISO 7730.

*Convection:* For the convective heat transfer, we use

$$\dot{Q}_{\mathrm{h,rh}} = h_{\mathrm{rh}} \cdot A_{\mathrm{rh}} \cdot (T_{\mathrm{rh}} - T_{\mathrm{cab}})\,, \tag{2}$$

$$\dot{Q}_{\mathrm{h,s,i}} = h_{\mathrm{s,i}} \cdot A_{\mathrm{body}} \cdot (T_{\mathrm{cab}} - T_{\mathrm{s,i}})\,, \tag{3}$$

$$\dot{Q}_{\mathrm{h,s,o}} = h_{\mathrm{s,o}} \cdot A_{\mathrm{body}} \cdot (T_{\mathrm{s,o}} - T_{\infty})\,, \tag{4}$$

for the surfaces of the RHs and for the inside and outside shell, respectively. The values for the convective heat transfer coefficients $h_{\mathrm{rh}} = 2.1\,\frac{\mathrm{W}}{\mathrm{m^2K}}$, $h_{\mathrm{s,i}} = 8.0\,\frac{\mathrm{W}}{\mathrm{m^2K}}$, and $h_{\mathrm{s,o}} = 20.7\,\frac{\mathrm{W}}{\mathrm{m^2K}}$ are based on EN ISO 6946 and are adjusted for an increased air movement inside the cabin and an average bus velocity of $15\,\frac{\mathrm{km}}{\mathrm{h}}$. The surface area of the entire bus hull is represented by $A_{\mathrm{body}} = 199\,\mathrm{m^2}$. The variable $A_{\mathrm{rh}}$ denotes the RH surface area, which is varied among simulations.

*Conduction:* For the conductive heat transfer through the shell, we use

$$\dot{Q}_{\mathrm{k}} = k_{\mathrm{body}} \cdot A_{\mathrm{body}} \cdot (T_{\mathrm{s,i}} - T_{\mathrm{s,o}})\,, \tag{5}$$

with the mean heat conductance through the bus enclosure given by $k_{\mathrm{body}} = 6.9\,\frac{\mathrm{W}}{\mathrm{m^2K}}$. This value is calculated to match an overall U-value of $2.9\,\frac{\mathrm{W}}{\mathrm{m^2K}}$, which has been determined for a city bus in a climate chamber by Sidler et al. (2019). As this U-value is more than ten times larger than the one of modern buildings, very high heat losses through the shell of the bus are to be expected.

*Door Losses:* To estimate the losses through open doors, we rely on the models developed by Schälin (1998), which are based on Bernoulli's law and the ideal gas law. Accordingly, we derive the following equation for the heat loss through the doors with heights $h_{\text{door}} = 1.95\,\text{m}$ and a combined overall width $w_{\text{door,tot}} = 4.4\,\text{m}$:

$$\dot{Q}_{\text{door}} = \frac{\rho_\infty \cdot c_{\text{p,a}} \cdot C_{\text{d}} \cdot \sqrt{g \cdot h_{\text{door}}^3}}{3} \cdot \sqrt{\frac{|T_{\text{cab}} - T_\infty|}{T_\infty}} \cdot (T_{\text{cab}} - T_\infty) \cdot w_{\text{door,tot}} \cdot \zeta_{\text{door}} \,. \tag{6}$$

The variables $\rho_\infty = 1.25\,\frac{\text{kg}}{\text{m}^3}$ and $c_{\text{p,a}} = 1005\,\frac{\text{J}}{\text{kg\,K}}$ represent the air density and heat capacity, respectively, while $g = 9.81\,\frac{\text{m}}{\text{s}^2}$ represents the gravitational acceleration. The discharge coefficient $C_{\text{d}} = 0.6$ is an empirical factor accounting for various sources of losses in the flow. The fraction $\zeta_{\text{door}} \in [0, 1]$ allows to specify the fraction of time the doors are open.

*Solar Irradiation:* We consider both direct and diffuse solar irradiation. For this purpose, we use measured data of the direct normal irradiance (DNI) $I_{\text{dni}}$ and the diffuse horizontal irradiance (DHI) $I_{\text{dhi}}$ obtained from a meteorological station of the Swiss Federal Office of Meteorology and Climatology (MeteoSwiss). Assuming a solar altitude angle $\beta$, a solar azimuth angle $\phi$, and isotropic diffuse radiation while neglecting ground-reflected irradiance, the irradiance on the roof and the wall of the bus are given by

$$I_{\text{roof}} = \cos\left(\frac{\pi}{2} - \beta\right) \cdot I_{\text{dni}} + I_{\text{dhi}} \,, \tag{7}$$

$$I_{\text{wall}} = \cos\beta \cdot \max\{\cos(\phi - \psi),\ 0\} \cdot I_{\text{dni}} + \frac{1}{2} \cdot I_{\text{dhi}} \,, \tag{8}$$

respectively, where the surface azimuth $\psi$ corresponds to the azimuth angle of the wall surface normal. Assuming that the bus is driving in all directions with equal probability (i.e., averaging over $\psi$), the mean irradiance on the wall can be simplified to

$$\bar{I}_{\text{wall}} = \frac{\cos\beta}{\pi} \cdot I_{\text{dni}} + \frac{1}{2} \cdot I_{\text{dhi}} \,. \tag{9}$$

Hence, the total heat by solar irradiation incident on the outside shell of the cabin is given by

$$\dot{Q}_{\text{sol,s,o}} = (1 - \zeta_{\text{sh}}) \cdot \Big(A_{\text{roof}} \cdot I_{\text{roof}} \cdot \alpha_{\text{paint}} \cdot (1 - \zeta_{\text{roof}}) + A_{\text{wall}} \cdot \bar{I}_{\text{wall}} \cdot (1 - \zeta_{\text{win}}) \cdot \alpha_{\text{paint}}\Big) \,, \tag{10}$$

where $A_{\text{roof}} = 48.6\,\text{m}^2$ and $A_{\text{wall}} = 102\,\text{m}^2$ represent the roof surface area and the combined surface area of all walls. For all surfaces, which are painted in bright colors, we use $\alpha_{\text{paint}} = 0.3$ based on data published by Incropera et al. (2007). The fractions $\zeta_{\text{roof}} = 0.7$ and $\zeta_{\text{win}} = 0.35$ represent the fraction of the roof that is shaded by components such as batteries, and the fraction of the walls that are windows, respectively. For the fraction $\zeta_{\text{sh}}$ representing the fraction of time where the bus is in the shade (e.g., by buildings or bridges), we use season-dependent values for urban areas as suggested by Centeno Brito et al. (2021).

The solar radiation transmitted into the bus through the windows is assumed to be fully absorbed by the interior of the bus. A fraction $\zeta_{\text{cab}} = 0.5$ of this heat denoted by $\dot{Q}_{\text{sol,cab}}$ is assumed to be absorbed directly by the cabin air, which allows to represent surfaces that are not directly connected to the shell, such as seats or passenger surfaces. The remainder of the incident irradiation, denoted by $\dot{Q}_{\text{sol,s,i}}$, is absorbed by the inner shell:

$$\dot{Q}_{\text{sol,cab}} = (1 - \zeta_{\text{sh}}) \cdot A_{\text{wall}} \cdot \bar{I}_{\text{wall}} \cdot \zeta_{\text{win}} \cdot \tau_{\text{win}} \cdot \zeta_{\text{cab}} \,, \tag{11}$$

$$\dot{Q}_{\text{sol,s,i}} = (1 - \zeta_{\text{sh}}) \cdot A_{\text{wall}} \cdot \bar{I}_{\text{wall}} \cdot \zeta_{\text{win}} \cdot \tau_{\text{win}} \cdot (1 - \zeta_{\text{cab}}) \,, \tag{12}$$

where the transmissivity of the windows to solar radiation is given by $\tau_{\text{win}} = 0.8$ based on data published by Incropera et al. (2007).

*Thermal Radiation:* For simplicity, all surfaces are assumed to be perfect emitters, which according to Incropera et al. (2007) does not introduce any major errors at room temperature. The windows are assumed to be opaque to this far-infrared radiation. Assuming a uniform ambient temperature, the heat incident on the outside of the bus is thus given by

$$\dot{Q}_{\text{r,s,o}} = \sigma \cdot A_{\text{body}} \cdot (T_{\text{s,o}}^4 - T_\infty^4) \,, \tag{13}$$

where $\sigma = 5.67 \times 10^{-8}\,\frac{\text{W}}{\text{m}^2\text{K}^4}$ is the Stefan-Boltzmann constant. Assuming that all RH panels are mounted in a coplanar arrangement on the ceiling of the bus, their mutual view factor is zero and thus, the view factor from the infrared heaters to the inner bus shell is one. Thus, the net radiative heat transfer from the RHs to the inside shell becomes

$$\dot{Q}_{\text{r,rh}} = \sigma \cdot A_{\text{rh}} \cdot (T_{\text{rh}}^4 - T_{\text{s,i}}^4) \,. \tag{14}$$

*Controlled Heat Flows:* The power necessary for heating the RH panels is given by $P_{\text{rh}}$. We assume that this electric power is fully converted into heat and that the RH panels are insulated against the ceiling. The heat provided by the HVAC system is denoted by $\dot{Q}_{\text{hvac}}$. Utilizing a vapor compression cycle (VCC) allowing air-conditioning (AC) and HP modes, the electric power can be calculated by

$$P_{\text{hvac}} = \frac{|\dot{Q}_{\text{hvac}}|}{\gamma} \,, \tag{15}$$

where $\gamma = \gamma(T_{\text{cab}} - T_\infty)$ represents its coefficient of performance (COP) as a function of the temperature difference between the two thermal reservoirs for both heating and cooling. This simple function is fitted to measurement data furnished by the VCC manufacturer of various operating points and achieves an $R^2$ value of more than 90%. It is shown in the top left graph of Fig. 2. Purely resistive heating, such as heating based on based on positive temperature coefficient (PTC) elements, can also be considered by selecting a heating COP of one.

*Heat Balance:* In our experience gained from climate chamber experiments, which is supported by Riachi and Clodic (2014) or Liebers et al. (2017), for instance, the time constant of the thermal dynamics of city buses is in the order of tens of minutes. On the other hand, the relevant disturbances either have time constants in the order of tens of seconds, as for instance in the case of door openings and the driving profile, or of hours, as in the case of the ambient temperature. Hence, for high-level energetic analyses, the operation of the HVAC system of a city bus can be approximated by the system's steady state.

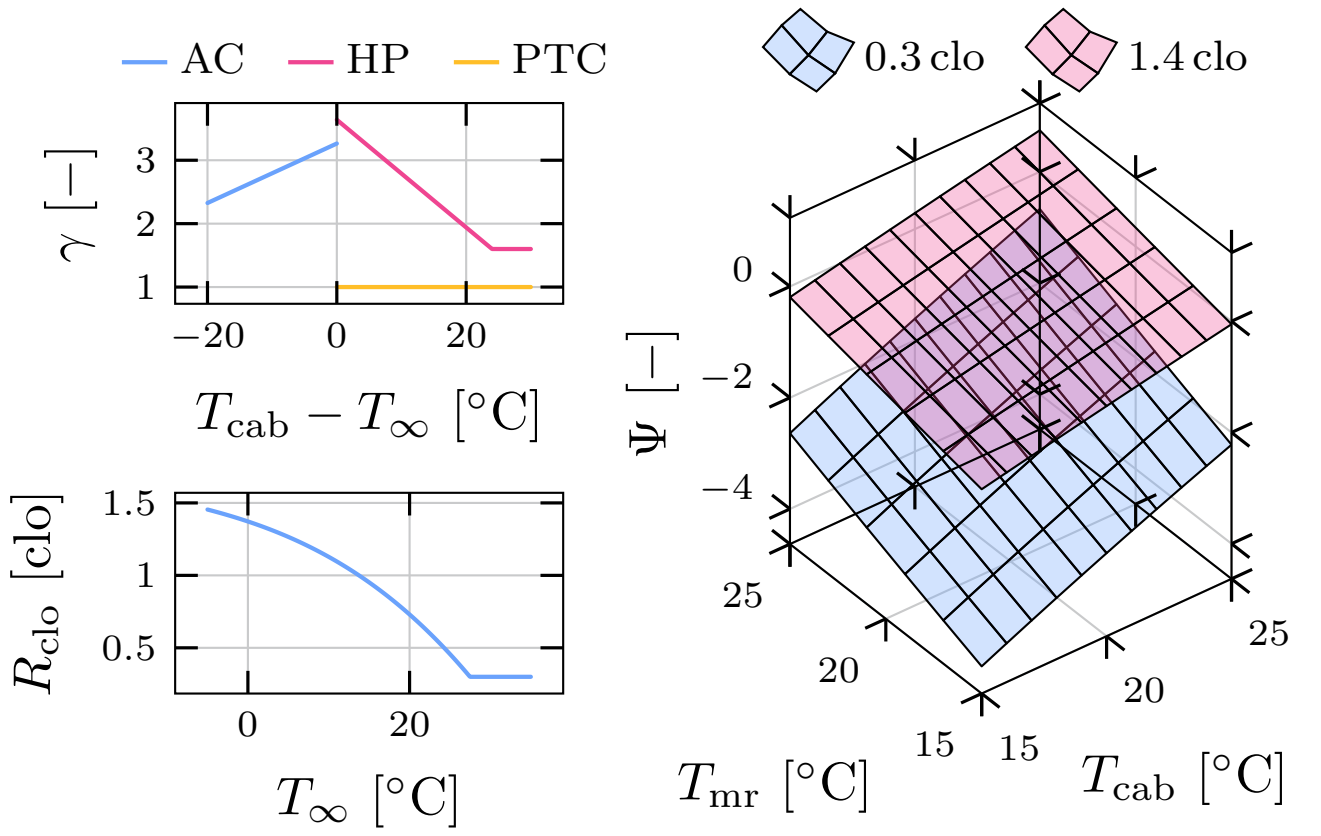

Fig. 2. Temperature-dependent model visualizations. The left graphs show the COP of the VCC and the clothing insulation. The right graph shows the predicted mean vote (PMV) for two different clothing values.

The following system of equations thus can be formulated for the heat balance:

$$0 = \begin{bmatrix} \dot{Q}_{\mathrm{pass}} + \dot{Q}_{\mathrm{h,rh}} - \dot{Q}_{\mathrm{h,s,i}} - \dot{Q}_{\mathrm{door}} + \dot{Q}_{\mathrm{sol,cab}} + \dot{Q}_{\mathrm{hvac}} \\ P_{\mathrm{rh}} - \dot{Q}_{\mathrm{r,rh}} - \dot{Q}_{\mathrm{h,rh}} \\ \dot{Q}_{\mathrm{r,rh}} + \dot{Q}_{\mathrm{h,s,i}} + \dot{Q}_{\mathrm{sol,s,i}} - \dot{Q}_{\mathrm{k}} \\ \dot{Q}_{\mathrm{k}} - \dot{Q}_{\mathrm{h,s,o}} - \dot{Q}_{\mathrm{r,s,o}} + \dot{Q}_{\mathrm{sol,s,o}} \end{bmatrix}, \tag{16}$$

where the four equations correspond to the energy conservation equations of the respective thermal reservoirs.

### 2.2 Thermal Comfort

As originally proposed by Fanger (1970), we model thermal comfort according to the norm EN ISO 7730, which characterizes the thermal comfort in terms of the PMV, ranging from −3 ("cold") to +3 ("hot"). For our calculations, we assume a constant air velocity of $v_{\mathrm{cab}} = 0.1\,\frac{\mathrm{m}}{\mathrm{s}}$ and a constant relative humidity of $\phi_{\mathrm{cab}} = 40\%$. For the metabolic heat rate, we use 1.2 met, which is consistent with the assumption used in Eq. (1). Thus, the PMV can be calculated as a function of the air temperature $T_{\mathrm{cab}}$, the mean radiant temperature $T_{\mathrm{mr}}$, and the clothing insulation factor $R_{\mathrm{clo}}$ as follows,

$$\Psi = f(T_{\mathrm{cab}}, T_{\mathrm{mr}}, R_{\mathrm{clo}}(T_\infty), v_{\mathrm{cab}}, \phi_{\mathrm{cab}}, \dot{Q}_{\mathrm{pass}}), \tag{17}$$

for which we show two sample visualizations in the right graph of Fig. 2. The insulation factor $R_{\mathrm{clo}}$ quantifies the clothing of the passengers, which depends on the ambient temperature as shown in the bottom left graph of Fig. 2. The relationship we use for this purpose is based on a model presented by Havenith et al. (2012). In accordance with examples from EN ISO 7730, we additionally introduce a lower limit of $R_{\mathrm{clo}} = 0.3$ in summer since we do not expect the clothing of passengers to be lighter.

To determine the mean radiant temperature $T_{\mathrm{mr}}$, we describe a passenger by an upright cuboid with the dimensions $0.25\,\mathrm{m} \times 0.25\,\mathrm{m} \times 1.7\,\mathrm{m}$, which does not take part in the radiation balance of the enclosure. The passenger can thus be considered a "passive" sensory device, apart from its metabolic heat release described by Eq. (1). Using the view factor relations given in Incropera et al. (2007), the mean radiant temperature of a passenger is represented by

$$T_{\mathrm{mr}}^4 = \frac{T_{\mathrm{s,i}}^4 \cdot \sum_i (A_i \cdot F_{i\to \mathrm{s,i}}) + T_{\mathrm{rh}}^4 \cdot \sum_i (A_i \cdot F_{i\to \mathrm{rh}})}{\sum_i A_i}, \tag{18}$$

where the index $i$ is used to iterate over the five surfaces of the cuboid. The view factors necessary for this calculation can be determined analytically as shown by Gross et al. (1981). Finally, to obtain an overall estimate for the PMV for multiple passengers, we average the PMV values calculated in Eq. (17).

## 3. SOLUTION APPROACH

Our goal is to minimize the power demand

$$P_{\mathrm{tot}} = P_{\mathrm{rh}} + P_{\mathrm{hvac}}, \tag{19}$$

while restricting the thermal comfort to a certain window

$$\Psi \in [\Psi_{\mathrm{min}}, \Psi_{\mathrm{max}}]. \tag{20}$$

By varying the size of the PMV window, we can evaluate the trade-off between the power consumption and the comfort attained.

In the following, we present two solution approaches for determining the lowest possible power consumption for a specific scenario.

### 3.1 Optimization-Based Solution

In the first approach, we evaluate the model based on a formulation of an optimization problem. For this purpose, we reformulate the bidirectional heat provided by the HVAC system as

$$\dot{Q}_{\mathrm{hvac}} = \dot{Q}_{\mathrm{hp}} - \dot{Q}_{\mathrm{ac}}, \tag{21}$$

where $\dot{Q}_{\mathrm{hp}} > 0$ represents the heat provided by the heating system and $\dot{Q}_{\mathrm{ac}} > 0$ represents the heat removal by the AC system. Hence, the HVAC power demand can be reformulated as follows:

$$P_{\mathrm{hvac}} = \frac{\dot{Q}_{\mathrm{hp}}}{\gamma_{\mathrm{hp}}(T_{\mathrm{cab}} - T_\infty)} + \frac{\dot{Q}_{\mathrm{ac}}}{\gamma_{\mathrm{ac}}(T_{\mathrm{cab}} - T_\infty)}, \tag{22}$$

where $\gamma_{\mathrm{hp}}$ and $\gamma_{\mathrm{ac}}$ represent the COP for heating and cooling as a function of the temperature difference, respectively. In contrast to Eq. (15), this reformulation no longer relies on an absolute value and is therefore preferable for derivative-based optimization algorithms.

Additionally, the nonlinear relationship of Eq. (17), which is calculated in EN ISO 7730 in an iterative fashion, is approximated using a polynomial fit. Finally, we add a constraint to ensure that the RHs are operated in accordance with their specified temperature $T_{\mathrm{rh,tgt}}$:

$$T_{\mathrm{rh}} = T_{\mathrm{rh,tgt}}. \tag{23}$$

The resulting optimization problem reads as follows:

$$\begin{aligned} \underset{\dot{Q}_{\mathrm{hp}}, \dot{Q}_{\mathrm{ac}}, P_{\mathrm{rh}}}{\text{minimize}} \quad & P_{\mathrm{tot}}, \\ \text{s.t.} \quad & \text{Eqs. (16) and (20)–(23)}. \end{aligned} \tag{24}$$

Although in principle the reformulation introduced in Eqs. (21) and (22) allows simultaneous heating and cooling, the optimal solution does not exhibit such a suboptimal behavior.

The constraint shown in Eq. (23) corresponds to the RHs being always turned "on". However, in summer conditions, using the RHs obviously is not useful. Instead of including a binary decision in the optimization, we formulate a second optimization problem where the RHs are turned “off”, i.e., the corresponding equations are removed from the model. We then calculate the solutions to both optimization problems and select the solution with the lower overall power consumption $P_{\mathrm{tot}}$. Hence, we assume that the RHs are used only if they reduce the overall power consumption. This solution procedure implemented in Matlab takes about 0.2 s on a standard personal computer.

### *3.2 Alternative Approach*

If the solution is to be computed without the use of an optimization algorithm, we suggest an alternative approach based on a root-finding problem. For this purpose, we introduce a constraint for a fixed PMV target value $\Psi_{\mathrm{tgt}}$:

$$\Psi = \Psi_{\mathrm{tgt}}\,. \tag{25}$$

Together with Eqs. (16) and (23), we obtain a system of six equations, which can be solved for the six unknown variables $T_{\mathrm{cab}}$, $T_{\mathrm{rh}}$, $T_{\mathrm{s,i}}$, $T_{\mathrm{s,o}}$, $\dot{Q}_{\mathrm{hvac}}$, and $P_{\mathrm{rh}}$. Our approach therefore allows us to “invert the causality” by prescribing a certain PMV and solving for the heat inputs. Equivalently, the system without RHs can be completed with Eq. (25) and solved for the remaining four unknown variables. Similarly to the approach in Section 3.1, the solution with the lower overall power consumption $P_{\mathrm{tot}}$ is selected.

To find the solution minimizing the power consumption for an allowed PMV window instead of the constraint shown in Eq. (25), we first solve a reduced system with $\dot{Q}_{\mathrm{hvac}} = 0$ and without Eq. (25). If this solution violates Eq. (20), we solve it again by setting the target PMV $\Psi_{\mathrm{tgt}}$ to the violated limit. Our implementation of this solution procedure takes about 0.3 s on a standard personal computer. As comparisons of the results show that both solution algorithms yield the same results, we use the first approach for our studies presented below.

## 4. SCENARIOS

Our goal is to evaluate the model not only for a specific selection of environmental conditions, but for a large dataset of environmental conditions that are representative of a year-round operation of the city bus. Therefore, in this section, we first introduce the dataset that we use for these simulations. Second, we describe how passengers are placed in the bus cabin. Finally, we explain the averaging scheme used to aggregate the results of the entire data set.

### *4.1 Data Sources*

The number of passengers $N_{\mathrm{pass}}$ and the fraction of the door openings $\zeta_{\mathrm{door}}$ are obtained from data recorded on a trolley bus in operation on various bus routes in Zürich. This data is part of the publicly available dataset presented by Widmer et al. (2023). The mean ambient temperature $T_\infty$, the DNI $I_{\mathrm{dni}}$, and the DHI $I_{\mathrm{dhi}}$ are obtained from measurement data of MeteoSwiss. We average these time-resolved signals over operation periods of one hour to get scalar values for the corresponding variables. Figure 3 provides an overview of the disturbances in all scenarios.

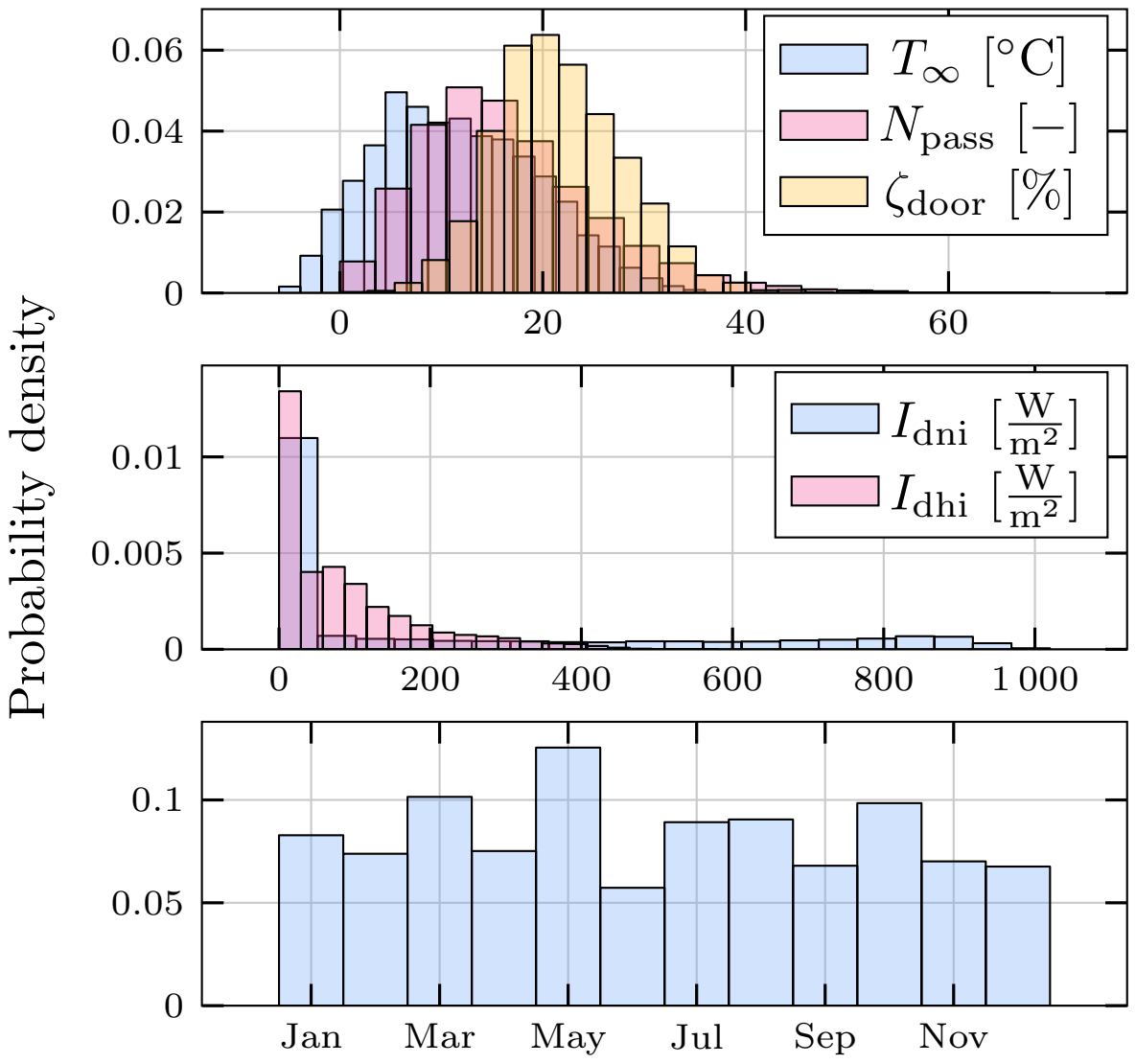

Fig. 3. Overview of the dataset of 7500 scenarios.

### *4.2 Passenger Placement*

The placement of the passengers in the bus influences their thermal comfort due to their relative position to the RH surfaces. Since passenger positions are not recorded during operation, we adopt a pseudo-random placement strategy that ensures a realistic distribution throughout the cabin. Figure 4 shows an example of such a distribution.

### *4.3 Data Aggregation*

As the scenarios recorded are not uniformly distributed throughout the year, as shown in Fig. 3, we average the results over each month, and then take the average of these values. In addition, comfort is maximized at a PMV value of zero, while higher or lower PMV values lead to reduced comfort levels. We therefore cannot average the resulting PMV values from scenarios of various climatic conditions. For instance, while the average of a PMV value of 1 in summer and −1 in winter is the same as the average of a PMV value of 0 across the whole year, the actual comfort level is better in the latter case. Therefore, we use the value for the predicted percentage dissatisfied (PPD), as suggested by EN ISO 7730. This quantity can be averaged over several scenarios without the issues mentioned above.

## 5. RESULTS

For illustration purposes, we first present results for single scenarios, followed by those that represent the year-round operation based on the averaging scheme introduced above.

### *5.1 Single-Scenario Results*

Figure 4 shows the mean radiant temperature at $N_{\mathrm{pass}} = 30$ different passenger locations, influenced by the RHs with a total surface area of $A_{\mathrm{rh}} = 4\,\mathrm{m}^2$ and a temperature $T_{\mathrm{rh,tgt}} = 70\,^\circ\mathrm{C}$. The RH layout chosen achieves a more or less uniform mean radiant temperature for most passengers, except for those close to the front or rear of the cabin, where the mean radiant temperature is close to the inside shell temperature of $T_{s,i} = 20\,°C$.

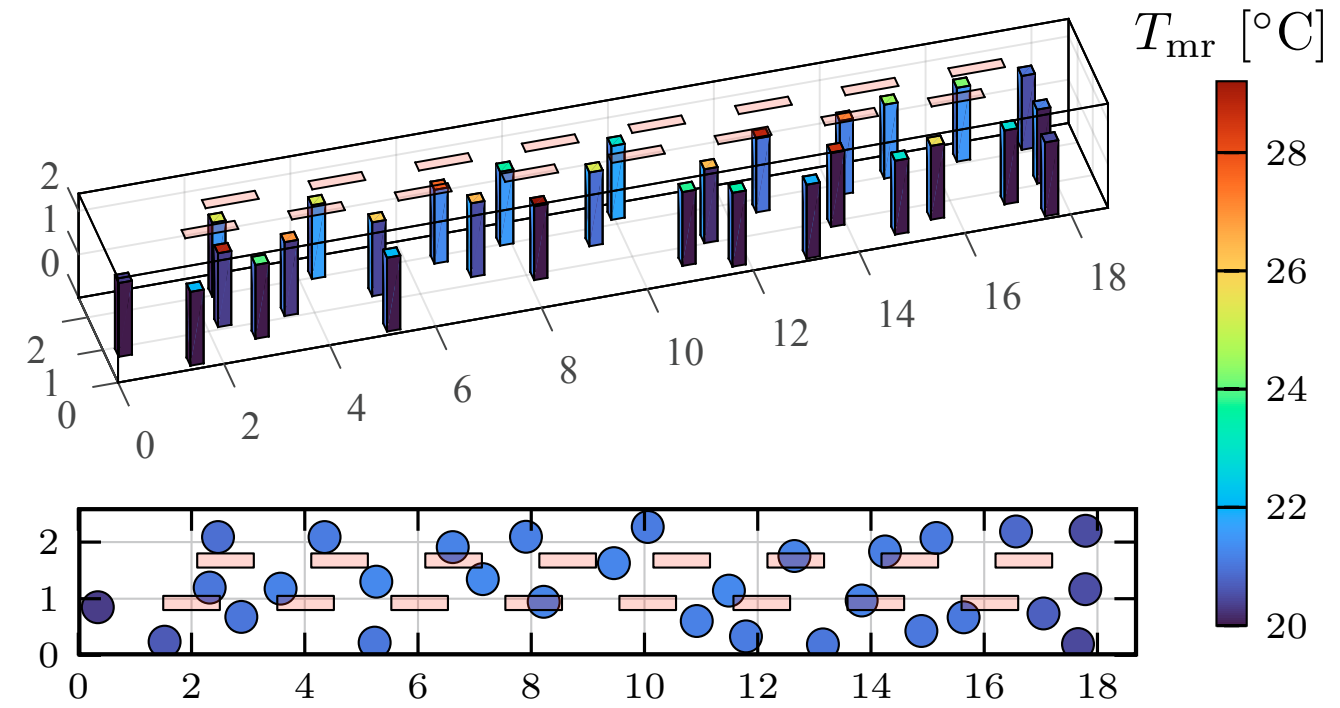

Fig. 4. Mean radiant temperatures perceived by passengers pseudo-randomly placed within the bus cabin (black outline) resulting from RH surfaces in the ceiling (light red). The isometric view on the top shows the mean radiant temperature of each surface, the top-view on the bottom shows each passenger's area-averaged mean radiant temperature.

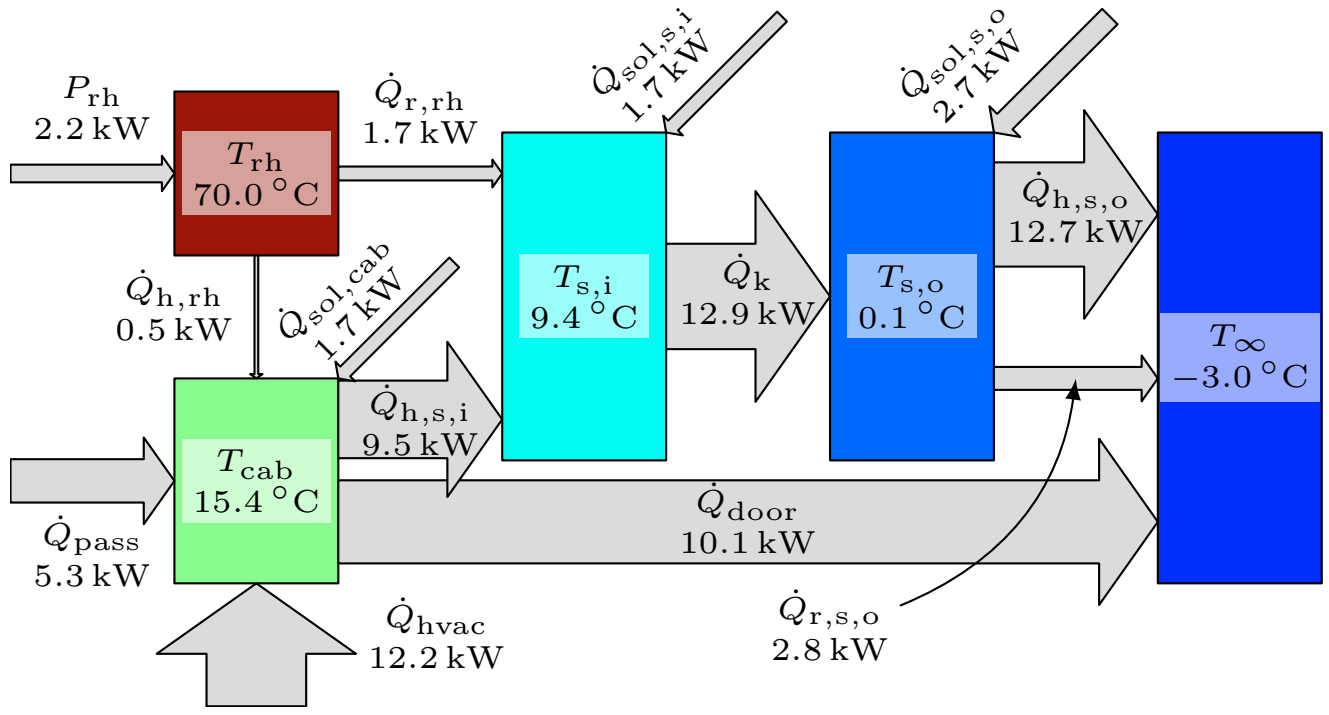

Fig. 5. Example heat flow overview for a winter scenario achieving "slightly cold" thermal comfort conditions, represented by $\Psi = -1$. The arrow annotations represent the corresponding absolute values.

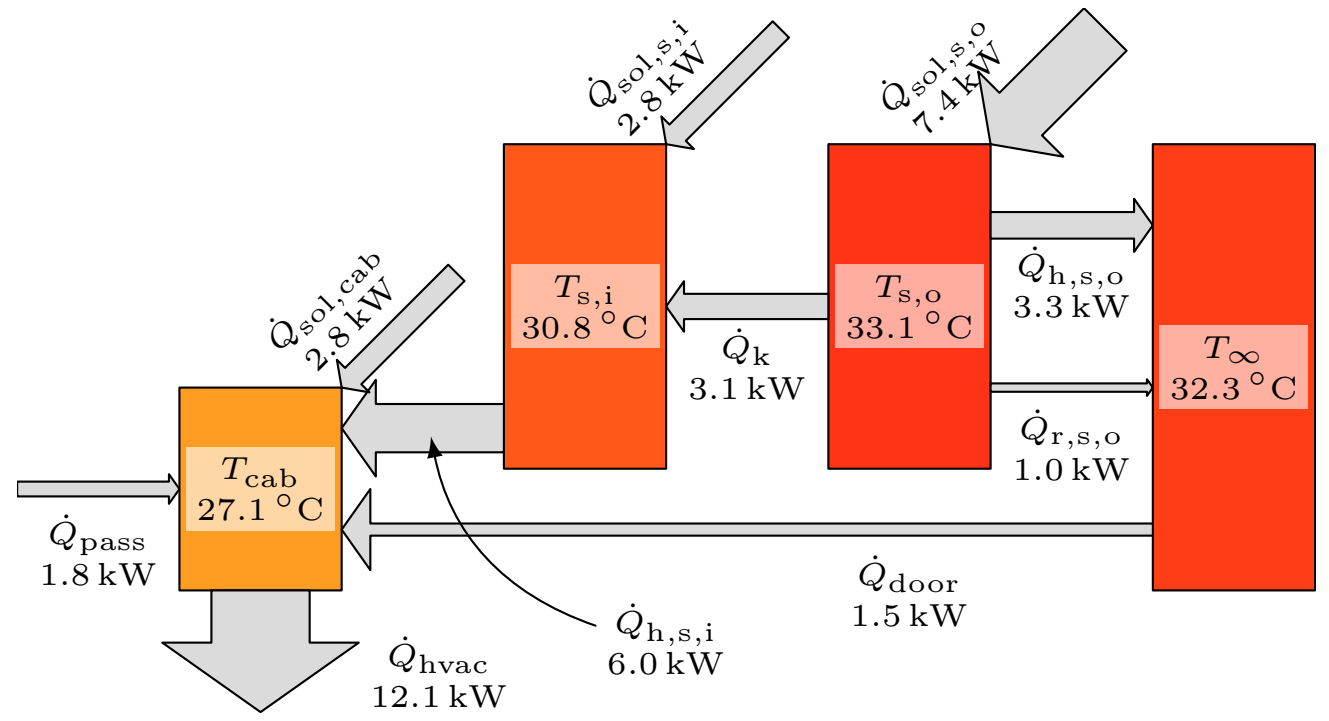

Fig. 6. Example heat flow overview for a summer scenario achieving "slightly warm" thermal comfort conditions, represented by $\Psi = 1$. The arrow annotations represent the corresponding absolute values.

Figures 5 and 6 visualize the optimal temperatures and power flows for a cold winter and a hot summer scenario, respectively. For the winter scenario, the optimal cabin air temperature is fairly low, which can be explained, on the one hand, by the high clothing insulation of $R_{clo} = 1.4$ clo and, on the other hand, by the ability of the RHs to improve the thermal comfort through an increased mean radiant temperature. Still, the temperature difference to the ambient is substantial, which drives the significant thermal losses through the shell and the doors. In typical summer scenarios, we note that the solar irradiation can be a major factor. On the other hand, the heat inputs through the shell and the doors typically is much less pronounced than in winter scenarios due to the smaller temperature difference between the cabin and the ambient.

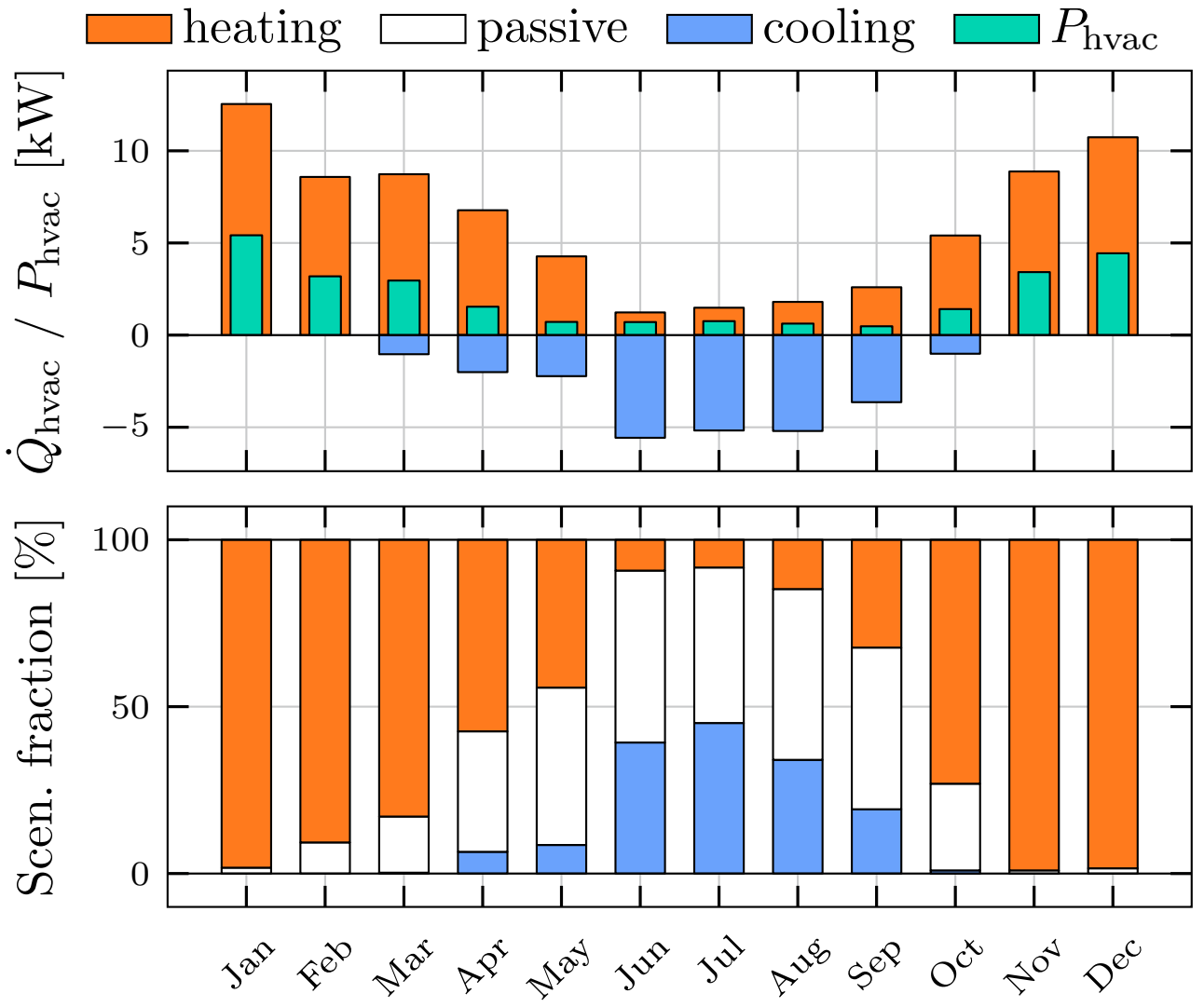

Fig. 7. Monthly averages of the heat $\dot{Q}_{hvac}$ and the power consumption $P_{hvac}$ required to achieve a PMV in the range $\Psi \in [-1, 1]$ in the upper graph and the fraction of the scenarios where heating or cooling is needed in the lower graph.

### 5.2 Year-Round Performance

*Monthly Averages:* Figure 7 visualizes the monthly averages of the heat request $\dot{Q}_{hvac}$ and the corresponding power consumption $P_{hvac}$. To simplify the interpretation of the results, no RHs are used in this case. The overview clearly shows that the average heat demand is greater in winter than in summer, which is to be expected for a central European climate. In addition, cooling is required in much fewer instances than heating. In fact, for this case study, the overall heating demand (positive values of $\dot{Q}_{hvac}$) is about 8 times larger than the cooling demand. In cases where the comfort requirements can be achieved without any heating or cooling, i.e., $\dot{Q}_{hvac} = 0$, the HVAC system is denoted "passive".

*Comparison of Heating Concepts:* As heating is the more influential case, we investigate the performance of different heating concepts in this section. On that account, we compare a system with a purely electrical PTC heater with an efficiency of 100%, i.e., $\gamma_{hp} = 1$, to a system with an R407c HP with a COP $\gamma_{hp}$, as shown in Fig. 2. We refer to the two HVAC systems as "PTC-AC" or "HP-AC", respectively. Both systems can be complemented by RHs. We do not change the cooling system, i.e., we always use the R407c AC system with a COP $\gamma_{ac}$ as shown in Fig. 2.

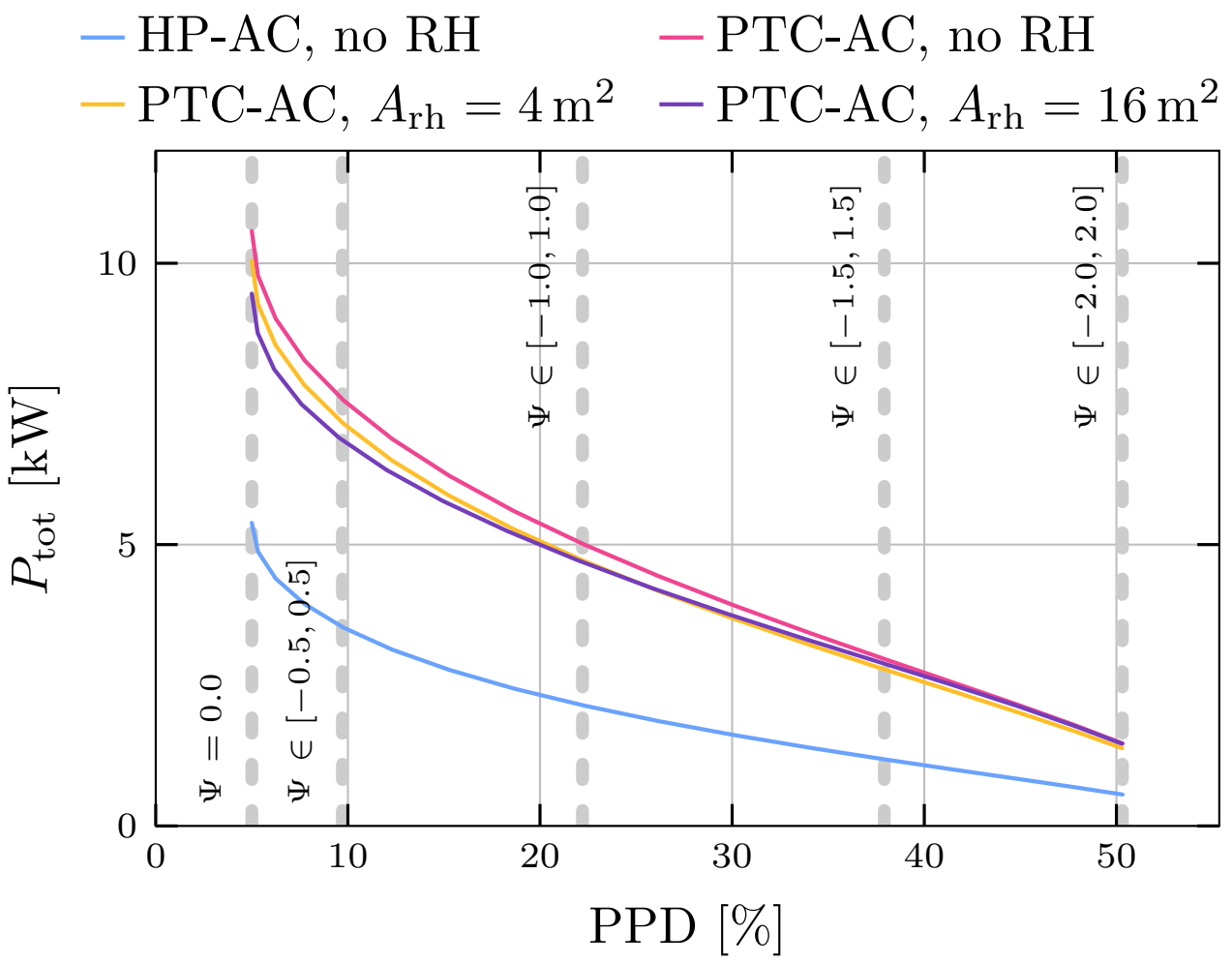

Fig. 8. Pareto front for different heating concepts showing the trade-off between the conflicting objectives of power demand and thermal discomfort. This analysis is based on all 7500 scenarios.

The trade-off between the power consumption and the thermal comfort achieved by the different systems throughout all scenarios is visualized in Fig. 8, where the allowed PMV is varied between $\Psi = 0$ (maximum comfort) and $\Psi \in [-2.0, 2.0]$ (least comfort). As explained in Section 4.3, we use the PPD value for data averaging. As expected from the definition of the PPD in EN ISO 7730, a value below 5% can never be reached in practice.

Figure 8 shows that RHs can reduce the mean HVAC power consumption of the PTC-AC system by 5% to 10% depending on the comfort requirements. Our analyses show that greater values of $T_{rh}$ generally improve the performance of the RHs due to the fourth power dependence in radiation compared to convection. Therefore, we use a very high value of $T_{rh,tgt} = 90\,°C$ for deriving the results shown in Fig. 8. Furthermore, Fig. 8 shows that increasing the RH panel size is beneficial for high comfort requirements. While a practical implementation of very large heater surfaces and high temperatures is questionable, the configuration serves as a best-case estimate of the energetic advantages achievable.

Comparing the PTC-AC heating system with an HP-AC system, Fig. 8 shows that the latter can reduce the consumption by 50% to 60%. An extensive search for RH configurations (size and temperature) to complement the HP reveals that the reduction in power consumption is limited to around 0.3%. Considering the added complexity of an RH installation and the costs of such an investment, RHs are not a reasonable addition to an HP. This observation was confirmed in simulations conducted with lower and higher ambient annual mean temperatures.

*Sensitivity Study:* The results of a one-at-a-time (OAT) sensitivity study are shown in Fig. 9. Similarly to the data shown in Fig. 7, this study is also conducted on an HP-AC system without RHs, based on the consideration of all 7500 scenarios. The metabolic heat rate $\dot{Q}_{met}$ has the highest relative influence on the mean power consumption. It influences the model in two ways: First, as a heat input in Eq. (1), and second, in the calculation of the PMV. Unsurprisingly, the clothing insulation factor $R_{clo}$ also has a strong influence on the mean power consumption. In terms of design variables, the largest leverage is expected by an increase in $\gamma_{hp}$. On the other hand, as expected due to the lower cooling requirements shown in Fig. 7, the advantage provided by an increase in $\gamma_{ac}$ is not nearly as large. This observation encourages the use of alternative refrigerants such as $CO_2$ (R744), which offer advantages in heating efficiency at the cost of lower cooling efficiency.

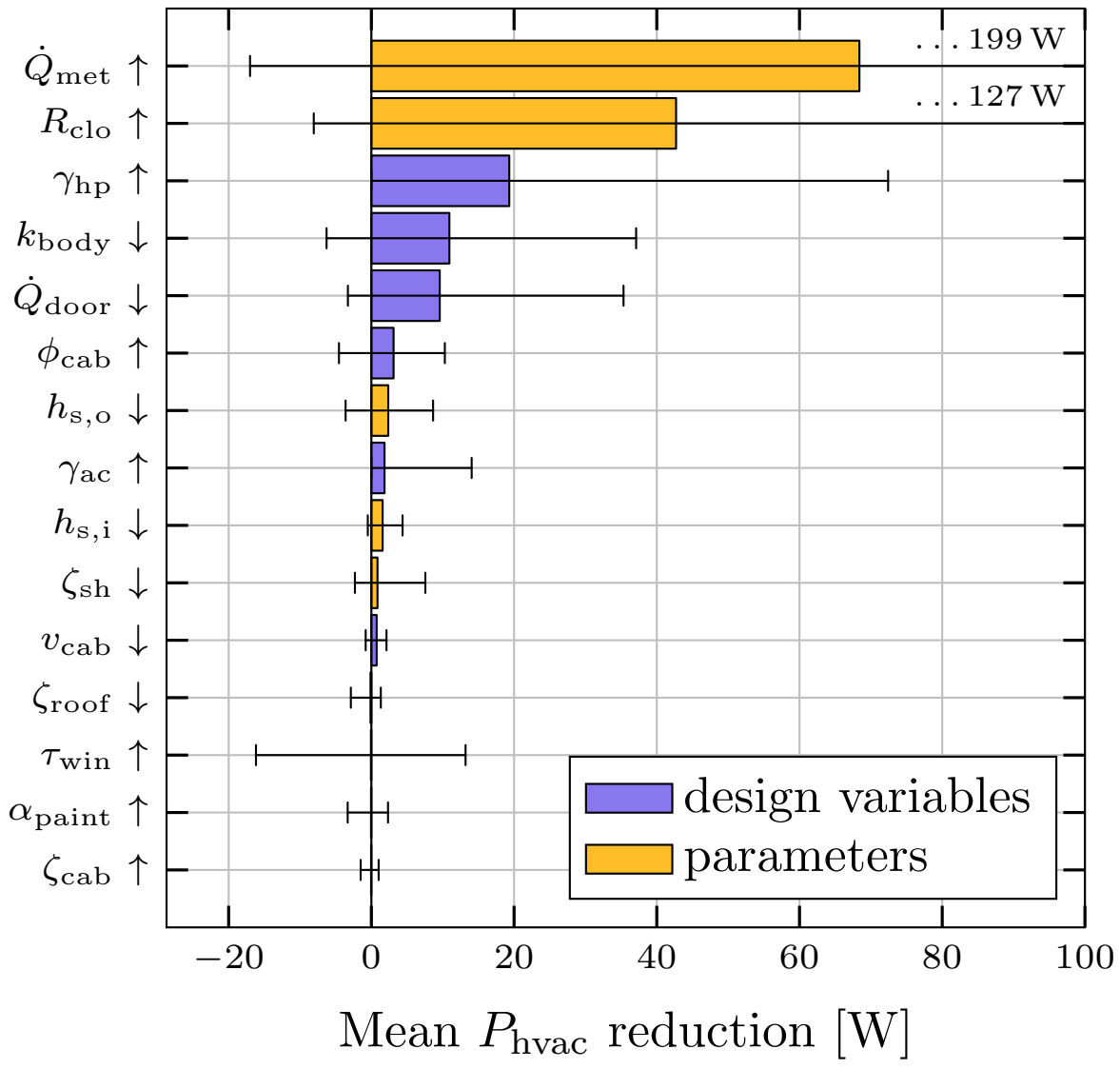

Fig. 9. Sensitivity of the annual mean power consumption with respect to 1% changes in model parameter values. Each parameter is marked by arrows ↑ or ↓ to indicate if its value is increased or decreased by 1%, respectively. To increase clarity, parameters that can be influenced in the vehicle design process are marked as "design variables". The whiskers show the 5th and 95th percentiles of all scenarios, respectively. As two of them exceed the graph, they are represented by numerical values.

Increasing the insulation, which causes a decrease of the thermal conductance $k_{body}$, or decreasing the door losses, e.g., by using air curtains, can improve the energy efficiency in a similar manner. For most of the parameters affecting the solar heat input, such as the window transmissivity $\tau_{win}$, the positive and negative effects roughly cancel each other out.

## 6. CONCLUSION AND OUTLOOK

In this paper, we propose mathematical models of HVAC systems of an electric city bus, which can be used to efficiently derive the Pareto front between power consumption and comfort requirements subject to a large set of scenarios that represent a year-round operation. We propose a steady-state formulation for which two alternative solution approaches are suggested. Compared to a system relying on PTC heaters and without affecting thermal comfort, the results show that an HP can lower the overall annual HVAC power consumption by 50% to 60%. Alternatively, RHs can be used to complement PTC heaters to reduce the power consumption by up to 10%. However, no significant advantage is found if an HP is complemented with RHs. Finally, a sensitivity study motivates the further promotion of $CO_2$ HPs due to their potentially higher heating COP. Further increases in efficiency can be expected by improvements in insulation and a reduction of door losses, whereas material properties related to solar irradiation, such as absorptivity or transmissivity, were found to have only a negligible influence.

In addition to the evaluation of the system design, our proposed method represents a promising approach to generate set points for HVAC controllers. For instance, a transport provider can select an admissible PMV range based on the data shown in Fig. 8 and extract the pairs of cabin and ambient temperature values from the corresponding simulation results. These pairs can serve as an informed basis to construct a reference temperature profile for the HVAC system, which currently is still determined on a heuristic basis for public transport vehicles.

For further research, we propose three distinct directions. First, the models used in this work could be extended to consider additional effects related to passenger comfort, such as humidity, air quality, ventilation-induced draft, radiant asymmetry, or heated seats. Second, using the proposed model for component sizing is a promising idea. For this purpose, the model needs to be extended to consider component limitations such as the maximum heating capacity on the one hand and the influence of the weights of the components on the energy consumption on the other. Third, additional case studies with different climatic conditions could be conducted to determine potential climatic tipping points among solutions that are favorable for heating scenarios such as $CO_2$ HPs, and ones that are favorable for cooling scenarios. In addition, the case studies could be extended to multiple types of public transport vehicles.